\documentclass[reprint,reprintnumbers,amsmath,amssymb]{revtex4-1}

\usepackage{graphicx}
\usepackage{dcolumn}
\usepackage{bm}

\usepackage{epstopdf} 
\newcommand{\D}{\text{D}}
\newcommand{\G}{\text{G}}
\newcommand{\T}{\text{T}}
\newcommand{\aIGZO}{\textit{a}-IGZO }

\begin{document}

\title{Competing weak localization and weak antilocalization in amorphous indium--gallium--zinc-oxide thin-film transistors}

\author{Wei-Hsiang Wang}
\affiliation{Department of Physics, National Taiwan Normal University, Taipei 116, Taiwan}
\author{Syue-Ru Lyu}
\affiliation{Department of Physics, National Taiwan Normal University, Taipei 116, Taiwan}
\author{Elica Heredia}
\affiliation{Department of Physics, National Taiwan Normal University, Taipei 116, Taiwan}
\author{Shu-Hao Liu}
\affiliation{Department of Physics, National Taiwan Normal University, Taipei 116, Taiwan}
\author{Pei-hsun Jiang}
\altaffiliation{E-mail: pjiang@ntnu.edu.tw}
\affiliation{Department of Physics, National Taiwan Normal University, Taipei 116, Taiwan}
\author{Po-Yung Liao}
\affiliation{Department of Physics, National Sun Yat-Sen University, Kaohsiung 804, Taiwan}
\author{Ting-Chang Chang}
\affiliation{Department of Physics, National Sun Yat-Sen University, Kaohsiung 804, Taiwan}
\author{Hua-Mao Chen}
\affiliation{Department of Photonics and Institute of Electro-Optical Engineering, National Chiao Tung University, Hsinchu 300, Taiwan}

\begin{abstract}
We have investigated the gate-voltage dependence and the temperature dependence of the magnetoconductivity of amorphous indium--gallium--zinc-oxide thin-film transistors. A weak-localization feature is observed at small magnetic fields on top of an overall negative magnetoconductivity at higher fields. An intriguing controllable competition between weak localization and weak antilocalization is observed by tuning  the gate voltage or varying the temperature. Our findings reflect controllable quantum interference competition in the electron systems in amorphous indium--gallium--zinc-oxide thin-film transistors.
\end{abstract}

\maketitle

Amorphous metal-oxide semiconductors have recently been studied for applications in thin-film transistors (TFTs) for large-area flexible electronics because of their electrical uniformity and fabrication advantage of room-temperature deposition and patterning \cite{Nomura04,Seo10,Gadre11,Chen09}. In particular, zinc oxide (ZnO) has recently attracted intense experimental and theoretical attention owing to its potential use in the emerging
nanoelectronics and optoelectronics \cite{Pearton03,Ozgur05,Klingshirn07}. ZnO-based semiconductors can incorporate indium oxide as a carrier-mobility enhancer and gallium oxide or hafnium oxide as  a columnar-structure suppressor for the amorphous phase in order to achieve high field-effect mobility and low off-state current of the channel \cite{Kim09,Kamiya10,Chung11}. They have become promising candidates of transparent and flexible nonvolatile memories to be integrated in system-on-panel displays \cite{Yabuta06,Yin08,Chen10}.

Aside from practical studies to achieve higher quality of amorphous indium--gallium--zinc-oxide (InGaZnO$_4$) TFTs, investigations of their fundamental electrical properties at low temperatures are necessary for studying quantum corrections to the conductivities of these carrier systems with disorders. Quantum interference and weak localization have been explored in three-dimensional and low-dimensional electron systems in various materials  \cite{Hikami80,Altshuler80,Ando82,Bergmann84,Abrahams01,Lee85,Lin02,Pierre03}. Indium zinc oxide (IZO) films and nanowires \cite{Shinozaki07,Shinozaki13,Thompson09,Chiu13,Kulbachinskii15,Xu10} in particular have raised special interest because of their potential applications in modern technologies. However, a comprehensive, in-depth study of low-temperature electrical transport in IZO is still missing, and the underlying mesoscopic and microscopic mechanisms remain largely unclear. Moreover, there are few detailed studies of low-temperature transport properties of practical IZO transistor devices. Measurements of IZO transistors at low temperatures may reveal interesting quantum-mechanical phenomena.

In this work, we present a study of the drain--source channel magnetoconductivity (MC) of an amorphous InGaZnO$_4$ (\textit{a}-IGZO) TFT measured at cryogenic temperatures. Manipulated via electric gating, the MC reveals a competition between weak localization (WL) and weak antilocalization (WAL) at small magnetic fields, where the WL component stays small but steady, while the WAL component lessens drastically with decreasing gate voltage. On the other hand, the temperature dependence of the MC also indicates a competition between WL and WAL, during which the WL component oscillates only marginally about a certain value, but the WAL component decreases gradually with increasing temperature. The competition between WL and WAL is analyzed with a quantum interference theory for two-dimensional (2D) carrier systems.

The \aIGZO TFTs based on an etch-stopper structure with passivation capping were fabricated on glass substrates. First, 300-nm-thick Ti/Al/Ti was deposited as gate electrodes by rf-magnetic sputtering, followed by a 300-nm-thick SiN$_x$ gate insulating layer using plasma-enhanced chemical vapor deposition. Next, 60-nm-thick \aIGZO was deposited by sputtering at room temperature to serve as the channel layer. After that, 200-nm-thick SiO$_x$ was deposited by sputtering as the etching stop layer. The source and drain electrodes were then formed by sputtering 300-nm-thick Ti/Al/Ti, followed by sputter deposition of 200-nm-thick SiO$_x$ as the passivation layer.

\begin{figure}
\includegraphics[width=3.5in]{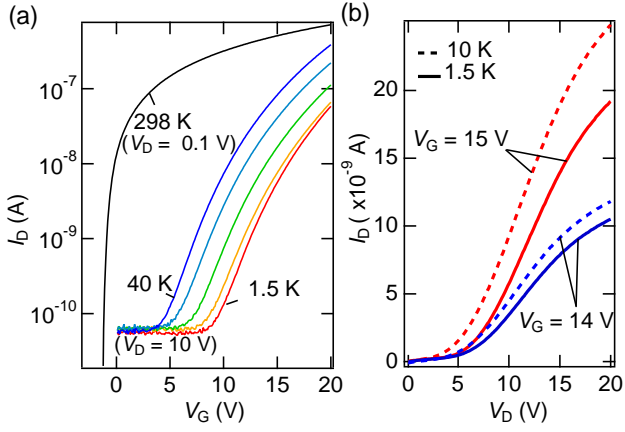}
\caption{\label{fig:fig1}
(a) Drain current ($I_\D$) of a \aIGZO TFT as functions of the gate voltage ($V_\G$) at various temperatures ($T$). The data traces from right to left were taken at 1.5, 10, 20, 30, and 40 K, respectively, with a fixed drain voltage $V_\D=10$ V, while the leftmost 298-K trace was taken with  $V_\D=0.1$ V. (b) Representative $I_\D$ vs $V_\D$ at different $V_\G$ and $T$.}
\end{figure}

In our experiment, we performed electrical measurements on several \aIGZO TFTs, and have observed similar behaviors from them at room and cryogenic temperatures ($T$). In this paper, we present the data collected from a representative \aIGZO TFT with a channel width ($W$) of 12 $\mu$m and a channel length ($L$) of 8 $\mu$m. Fig.~\ref{fig:fig1}a shows the drain-to-source current ($I_\D$) of the \aIGZO TFT as functions of the gate-to-source voltage ($V_\G$) with a fixed drain voltage ($V_\D$) of 10 V measured at various low $T$ from 1.5 K to 40 K, while the room-temperature trace also shown in the figure was measured at $V_\D=0.1$ V. The room-temperature electrical characteristics of the device include a fast switching time ($\sim$ ms), a threshold gate voltage ($V_\T$) of $-1.4$ V, a subthreshold swing (S.S. $= dV_\G/d(\log I_\D$) ) of 0.45 V/decade, and a mobility of 5.6 cm$^2$/Vs. While a small $V_\D$ was used to operate the device at room temperature, larger $V_\D$ was needed at cryogenic temperatures to ``unfreeze" the carrier transport in the channel. Shown in Fig.~\ref{fig:fig1}b are the  representative $I_\D$--$V_\D$ traces, from which one can notice that $V_\D=10$ V at either 1.5 K or 10 K drives the electrical transport away from its off-state to an intermediate regime far before its tendency towards saturation. In contrast to standard transport measurements where conductivities of thin metal films are obtained in the limit of zero driving voltage \cite{Dolan79},  moderate $V_\D$ is needed for \aIGZO TFTs at low $T$ to initiate sufficient $I_\D$ for valid conductivity measurements. From Fig.~\ref{fig:fig1}a, it can be seen that $V_\T$ shifts to higher values as the $T$ decreases, and S.S. is seriously enlarged as well.

\begin{figure}
\includegraphics[width=3.5in]{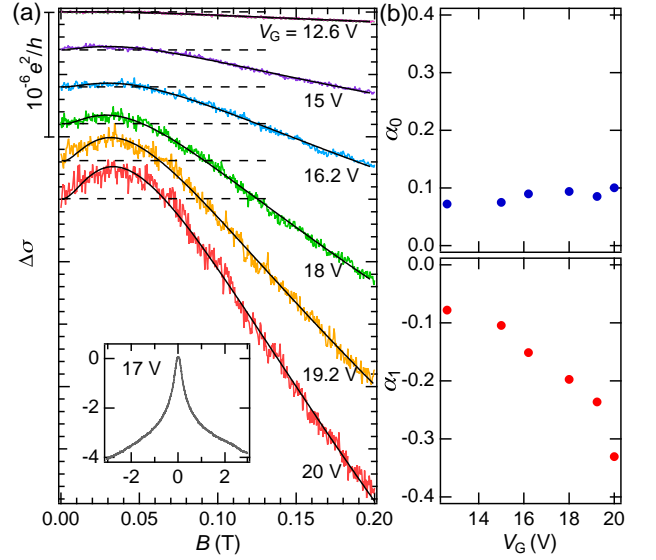}
\caption{\label{fig:fig2}
(a) Normalized magnetoconductivity ($\Delta \sigma$) of a \aIGZO TFT as functions of the magnetic field ($B$) at various fixed gate voltages ($V_\G$) from 12.6 to 20 V with $V_\D=10$ V and $T=1.5$ K. Traces are vertically offset for clarity with horizontal dashed lines indicating the corresponding zeros. Theoretical fits (Eq.~\ref{eq:1}) are shown as solid smooth lines. Inset: $\Delta \sigma$ vs $B$ at $V_\G=17$ V. (b) $\alpha_0$ and $\alpha_1$ vs $V_\G$ obtained from (a).}
\end{figure}

$I_\D$ vs $V_\G$ at low $T$ presented above implies a transport mechanism different from that at room temperature. In order to study the underlying physics, the magneto-transport properties of the \aIGZO TFT was investigated in the quantum diffusive regime. Shown in Fig.~\ref{fig:fig2}a is the normalized MC ($\Delta \sigma(B)=\sigma(B)-\sigma(0)$ )   obtained at various fixed $V_\G$ from 12.6 to 20 V as functions of the magnetic field ($B$) perpendicular to the channel plane, where the conductivity $\sigma$ is defined as $I_\D/V_\D\cdot L/W$, i.e., the 2D conductivity of the channel. The traces are vertically offset for clarity with horizontal dashed lines indicating the corresponding zeros. The measurement was conducted at $T=1.5$ K with a fixed $V_\D$ of 10 V. $\Delta \sigma(B)$ traces are presented only for positive $B$ below 0.2 T for a clearer view, while all of them are indeed symmetric about $B=0$, like the representative trace shown in the inset  over a larger range of $B$. At $V_\G=20$ V, $\Delta \sigma(B)$ slightly increases as $B$ increases for $B \lesssim 0.035$ T, exhibiting a WL behavior. At higher $B$, however, the trace curve bends over to WAL. These features in the $\Delta \sigma(B)$ curve weaken as $V_\G$ is decreased. When $V_\G$ goes below 15 V, the local minimum around $B=0$ is smeared into a plateau, while the decrease of $\Delta \sigma(B)$ under higher $B$ remains detectable. This implicates a $V_\G$-controlled quantum interference competition between WL and WAL in the \aIGZO TFT. 

To analyze the relative strengths of WL and WAL contributions as functions of $V_\G$, we introduce the two-component Hikami--Larkin--Nagaoka (HLN) theory for the MC of a 2D system \cite{Hikami80,Lu11,Lu11a}:
\begin{equation}\label{eq:1}
\Delta\sigma(B)=A\sum_{i=0, 1}\frac{\alpha_i e^2}{\pi h}\Bigg[\Psi \Bigg(\frac{\ell_B^2}{\ell_{\phi i}^2}+\frac{1}{2} \Bigg) - \ln \Bigg( \frac{\ell_B^2}{\ell_{\phi i}^2} \Bigg)\Bigg],
\end{equation}
where $\Psi$ is the digamma function, $\ell_B \equiv \hbar/(4e|B|)$ is the magnetic length, the prefactors $\alpha_0$ and $\alpha_1$ stand for the weights of WL and WAL contributions respectively, and $\ell_{\phi i}$ is the corresponding phase coherence length. $A$ is a coefficient we added to the original two-component HLN equation, which will be discussed later. In the limit of pure WAL, $\alpha_0 = 0$, and $\alpha_1 = -1/2$ for each band that carries a $\pi$ Berry phase \cite{McCann06}. In the limit of pure WL, $\alpha_1 = 0$, and $\alpha_0 = 1$ for a usual 2D system and 1/2 for a topological surface state \cite{Lu11}. To all the $V_\G$-dependent $\Delta \sigma(B)$ traces in Fig.~\ref{fig:fig2}a, Eq.~\ref{eq:1} provides excellent fits shown as solid smooth lines, and the evolutions of $\alpha_0$ and $\alpha_1$ obtained from the fits are summarized in Fig.~\ref{fig:fig2}b as functions of $V_\G$. When $V_\G$ is decreased from 20 to 12.6 V, $|\alpha_1|$ (the prefactor for WAL) decreases considerably from 0.33 to 0.08, while $\alpha_0$ (the prefactor for WL) stays between 0.10 and 0.07. This indicates a drastically decreasing WAL component accompanied with a small but relatively steady WL contribution as $V_\G$ is decreased. When $V_\G$ reaches below 12.6 V, $|\alpha_1|$ approaches $\alpha_0$, leading to the weakest $B$ dependence of the MC (the \textit{unitary}-like behavior).

A sufficiently small value of $10^{-4}$ has to be assigned to the coefficient $A$ in Eq.~\ref{eq:1} in the analysis above, because the measured MC's of our devices are only minuscule fractions of $e^2/h$. For $V_\G$ from 12.6 to 20 V at $T=1.5$ K, the \aIGZO TFT is operated between the \textit{off} and \textit{on} (saturation) states with low carrier densities. If the interpretation were carried out using the original HLN theory (with $A=1$), the system in the \aIGZO TFT would be regarded as a unitary class with strong magnetic scattering even for $V_\G=20$ V, leaving the subtle evolutions of the WL and WAL unexplained. In order to perform a systematic analysis, the fitting procedure was started on the $\Delta \sigma(B)$ trace of $V_\G=20$ V to find a reasonable fitting coefficient. The extremely small MC may be explained in terms of a seriously suppressed channel of conduction in the carrier system, whereas the WL and WAL phenomena can be preserved in this scenario. Specific modification of the theoretical model is needed to interpret the electrical transports in semiconductor transistors.

\begin{figure}
\includegraphics[width=3.6in]{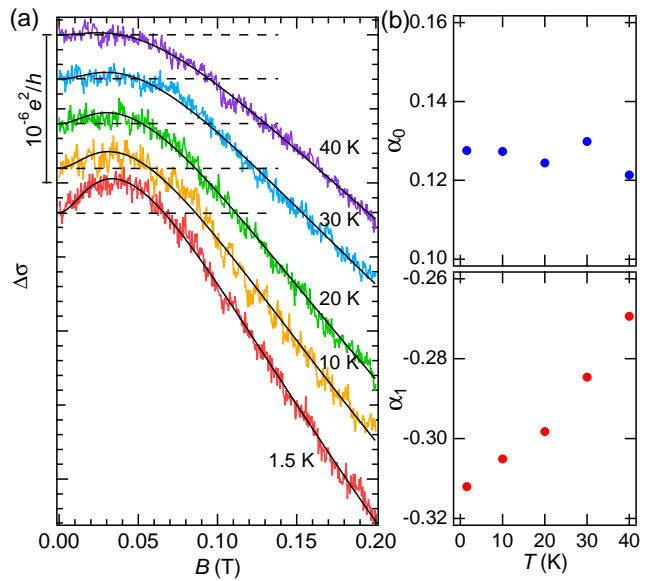}
\caption{\label{fig:fig3}
(a) Normalized magnetoconductivity ($\Delta \sigma$) of a \aIGZO TFT with $V_\D=10$ V as functions of the magnetic field ($B$) with $I_\D (B=0)$ held at $5\times 10^{-8}$ A by adjusting $V_\G$ at various temperatures ($T$) from 1.5 to 40 K. Traces are vertically offset for clarity with horizontal dashed lines indicating the corresponding zeros. Theoretical fits (Eq.~\ref{eq:1}) are shown as solid smooth lines. (b) $\alpha_0$ and $\alpha_1$ vs $T$ obtained from (a).}
\end{figure}

The $T$ dependence of the MC was also examined. Illustrated in Fig.~\ref{fig:fig3}a is $\Delta \sigma$ vs $B$ measured at various $T$ from 1.5 to 40 K. In this experiment, $I_\D$ at $B=0$ is held constant at $5\times 10^{-8}$ A for different $T$ from 1.5 to 40 K by adjusting $V_\G$ accordingly from 19.7 V to 13.9 V, through which the effect of the $T$-dependent $V_\T$ on transport can be compensated so that the transport conductivity at $B = 0$ is kept the same when the $T$ dependence of $\Delta \sigma(B)$ is being studied. (In contrast, the transport conductivity changes with $V_\G$ in the measurements shown in Fig.~\ref{fig:fig2}a.) One can find that the WL behavior  at $B \lesssim 0.035$ T gradually evolves into a plateau as $T$ is increased to 40 K, while the WAL behavior at higher $B$ weakens gradually with increasing $T$. At $T=40$ K (with $V_\G = 13.9$ V), $\Delta \sigma$ suffers a decrease of $\sim1.3 \times 10^{-6} e^2/h$ as $B$ is increased from 0 to 0.2 T, while $\Delta \sigma$ with similar $V_\G$ at 1.5 K exhibits a much smaller decrease of less than $\sim3 \times 10^{-7} e^2/h$, which can be read from Fig.~\ref{fig:fig2}a by examining the traces of $V_\G = 12.6$ V and 15 V. This distinguishes the $T$-dependent behaviors of  WL and WAL  from the $V_\G$-dependent behaviors observed in Fig.~\ref{fig:fig2}a. The data from the $T$-dependence measurements is also analyzed using Eq.~\ref{eq:1} with the results presented in  Fig.~\ref{fig:fig3}b. As $T$ is increased, $|\alpha_1|$ decreases gradually from 0.31 to 0.27, while $\alpha_0$ oscillates only slightly between 0.12 and 0.13.

Gate-voltage controlled weights of WL and WAL contributions demonstrated in Fig.~\ref{fig:fig2} are  an intriguing feature that has not been specifically reported for IZO thin films. Electric-field controlled competition of WL and WAL has been observed in ultra-thin topological insulators \cite{Lu11}, and is theoretically interpreted in terms of the Berry phase tuned by the Fermi level with respect to the gapped surface states \cite{Lang12}. A similar mechanism may explain the WL--WAL crossover in our \aIGZO TFTs, but it remains unclear how to translate the observed WL and WAL behaviors into the details of the band structure scanned via gate-voltage tuning. It requires further theoretical research to unveil the fundamental physics of the quantum interference in \aIGZO TFTs and related devices.

The $T$ dependence of the competition between WL and WAL shown in Fig.~\ref{fig:fig3} is also an important aspect in quantum interference study. A transition from a WAL to a WL state was found in three-dimensional \aIGZO films as $T$ is increased \cite{Shinozaki13}, while WAL in our device survives at higher $T$ with the WL behavior at $B \lesssim 0.035$ T evolves into a plateau. According to the HLN theory for 2D systems \cite{Hikami80}, the system favors WL when the energy relaxation time decreases at higher $T$ and  becomes shorter  than the spin--orbit interaction time or the magnetic scattering time. However, the persistent WL component in our device is found to be accompanied with a strong WAL component, although the contribution from the latter indeed decreases as $T$ is raised. More theoretical studies are needed to understand the underlying physics of the competing WL and WAL in \aIGZO TFTs.

\begin{figure}
\includegraphics[width=2.5in]{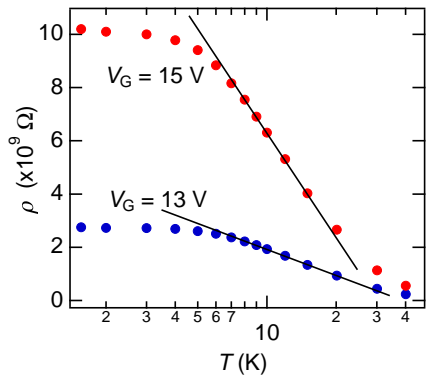}
\caption{\label{fig:fig4}
Resistivities ($\rho$) of a \aIGZO TFT with $V_\D = 10$ V as functions of the temperature ($T$) measured at $V_\G = 13$ V and  15 V, respectively. The guidelines of $-\log T$ dependence are shown as the solid lines.}
\end{figure}

The $T$ dependence of the zero-magnetic-field channel resistivity ($\rho$) of the \aIGZO TFT is displayed in Fig.~\ref{fig:fig4} with two representative data sets measured at $V_\G=15$ and 13 V, respectively, with $V_\D=10$ V. It can be seen that $\rho$ increases as $\Delta\rho \propto -\log T$ with decreasing $T$ from 20 K to 6 K. The $-\log T$ dependence of $\rho$ is reminiscent of the theoretical prediction for 2D systems with WL at low $T$. $\rho$ starts to deviate from the $-\log T$ guideline to a higher value as $T$ goes beyond 20 K, leaving the low-temperature regime where the theoretical prediction is assumed to hold. As $T$ is decreased below 6 K, on the other hand, $\rho$ no longer catches up the extrapolation of the $-\log T$ guideline. The lower end of the $T$ range over which the $-\log T$ dependence can hold is probably due to WAL. From Fig.~\ref{fig:fig3}b, it is clear that $|\alpha_1|$ (the WAL prefactor) is enhanced as $T$ is decreased. Therefore, $\rho$ values at lower $T$ deviate from the $-\log T$ dependence predicted for systems with only WL.

The validity of using a  2D-system theory to analyze the WL and WAL observed in our \aIGZO TFT may be inspected by considering the band energy of the \aIGZO channel \cite{Chen12}. As a positive $V_\G$ is applied to the device, the electrons are attracted to the interface of the \aIGZO active layer and the gate insulating layer, giving a tilted quantum well with an effective width of only a small fraction of the whole \aIGZO layer thickness (60 nm). Moreover, with an effective mass of 0.35 of the free-electron mass \cite{Abe12}, the spacing between the ground and the first excited state of the carrier system in the \aIGZO channel with $V_\G$ in the interested range is estimated to be on the order of 10--100 meV. Therefore, most of the carriers stay in their ground states at cryogenic temperatures, forming a quasi-2D  system.

In summary, we have measured the MC of an \aIGZO TFT at cryogenic temperatures down to 1.5 K. With increasing $V_\G$ from 12.6 to 20 V, the MC evolves from a unitary case to a nontrivial behavior, in which it slightly increases with increasing $B$ for $B \lesssim 0.035$ T, exhibiting a WL behavior, but then decreases into a WAL behavior at higher $B$. Analysis using the two-component HLN theory reveals a competition between WL and WAL, where the WL component stays small but steady, while the WAL component lessens drastically with decreasing $V_\G$. On the other hand, the $T$ dependence of the MC also indicates a competition between WL and WAL, during which the WL component oscillates only marginally about a certain value, but the WAL component decreases gradually with increasing $T$. The gate-voltage controlled competition between WL and WAL may be interpreted in terms of the phase change in quantum interference tuned by the Fermi level in the band structure, while the temperature-controlled behavior may be explained in terms of the change in the energy relaxation time. More theoretical studies are required to fully understand the underlying physics of the competing WL and WAL in \aIGZO TFTs.

The work was supported by the Ministry of Science and Technology of the Republic of China under Grant No. MOST 102-2112-M-003-009-MY3.

\end{document}